\documentclass[12pt,preprint]{aastex}

\begin{document}

\title{Precise Orbital Parameters and Anomalous Phase Variations of the Accretion-powered Millisecond Pulsar XTE J1807-294}

\author{Y. Chou, Y. Chung, C. P. Hu and T. C. Yang}
\affil{Graduate Institute of Astronomy, National Central University} 
\affil{No. 300, Jhongda Rd., Jhongli City, Taoyuan, Taiwan 320}
\email{yichou@astro.ncu.edu.tw}

\begin{abstract}
This study reports pulse variation analysis results for the forth discovered accretion-powered millisecond pulsar XTE J1807-294 during its 2003 outburst observed by {\it Rossi X-ray Timing Explorer}.  The pulsation is significantly detected only in the first $\sim$90d out of $\sim$150d observations.  The pulse phase variation is too complex to be described as an orbital motion plus a simple polynomial model.  The precise orbital parameters with $P_{orb}=40.073601(8)$ min and ${\it a_x}\sin {\it i}=4.823(5)$ lt-ms were obtained after applying the trend removal to the daily observed 150s segments pulse phases folded with a constant spin frequency without Keplerian orbit included.  The binary barycenter corrected pulse phases show smooth evolution and clear negative phase shifts coincident with the flares seen on the light curve and the enhancements of fractional pulse amplitude.  The non-flare pulse phases for the first $\sim$60d data are well described as a fourth order polynomial implying that the neutron star was spun-up during the first $\sim$60d  with a rate $\dot \nu=(1.7\pm0.3) \times 10^{-13}$ Hz/s at the beginning of the outburst.  Significant soft phase lags up to $\sim$500 $\mu s$ ($\sim$10\% cycle) between 2 to 20 keV were detected for the nonflare pulse phases. We conclude that the anomalous phase shifts are unlikely due to the accretion torque but could result from the ``hot spot'' moving on the surface of neutron star.

\end{abstract}

\keywords{accretion, accretion disks ---binaries: close --- pulsars: individual (XTE J1807-294) --- stars: neutron---X-rays: binaries}

\section{Introduction} \label{intro}
Low Mass X-ray Binaries (LMXBs) are considered to be the progenitor of the millisecond pulsars detected in the radio band (see \citealt{bha91} for an extensive review).  It is widely believed that the weakly magnetized, slowly rotating neutron star is gradually spun-up through the transfer of angular momentum carried by the matter from an accretion disk \citep{alp82,rad82}.  Millisecond-time-scale variations have been observed in many LMXB systems, including kilohertz Quasi-Periodic Oscillation \citep[kHz QPO, see][]{van05} and burst oscillation (see \citealt{str05} for a review); however, the attempts to measure coherent millisecond pulsation from an LMXB, which provides a direct evolutionary link between radio millisecond pulsar and LMXB, were unsuccessful until the discovery of the first accretion-powered millisecond pulsar SAX 1808.4-3658 with a spin period $P_{spin} \simeq 2.5$ ms \citep{wij98, cha98}, in 1998.  To date, coherent millisecond pulsations have been detected in ten LMXBs with spin periods ranging from 1.67 ms to 5.5 ms \citep{wij07,mar07,gav07,cas07,alt07}.

The fourth known accretion-powered millisecond pulsar, XTE J1807-294, was discovered on February 21, 2003 while the Proportional Counter Array (PCA) onboard the {\it Rossi X-ray Timing Explorer} ({\it RXTE}) monitored the Galactic-center region \citep{mar03a}. Its preliminary spin frequency of 190.625 Hz \citep{mar03a} and orbital period of $40.0741 \pm 0.0005$ min \citep{mar03b} were immediately revealed right after its discovery.  Follow-up observations were made with many X-ray observatories. \citet{mar03b} determined the source location, R.A.=$18^h06^m59^s.8$ DEC=$-29^{\fdg}24^{\arcmin}20^{\arcsec}$ (equinox 2000, $\sim1^{\arcsec}$ error) by using  {\it Chandra} observation. \citet{cam03}, \citet{kir04} and \citet{fal05a} reported more precise orbital parameters and pulsation period using {\it XMM-Newton} observations. However, \citet{mar04} found wide swings in the apparent spin change rate vs. accretion rate, which were interpreted as the hot spot moving on the neutron star rather than from accretion torque.  The wide-band (0.5 to 200 keV) spectrum obtained by the combination of simultaneous observations from {\it XMM-Newton}, {\it RXTE} and {\it INTEGRAL} is well described by an absorbed disk blackbody plus thermal Comptonization model \citep{fal05a}.  {\it RXTE} has processed extensive monitoring of XTE J1807-294 since February 27, 2003 with a time span of more than 150d until the source went into its quiescent state.  The light curve exhibited an exponential-like decay with a time-scale of $\sim 19$d at the beginning of the outburst followed by a much slower decay with a time constant of $\sim 122$d \citep{fal05a}.  Several broad soft flares lasting for hours to days were seen during the flux decay.  The pulsed fraction increased and the pulse profile became more sinusoidal during the flares \citep{zha06}.  Twin kHz QPOs were also detected in XTE J1807-294 by {\it RXTE} \citep{lin05}.

Owing to their fast rotation, the related parameters of the accretion-powered millisecond pulsars, such as orbital and spin parameters, can be precisely measured during their outbursts lasting for tens to hundreds of days. Unfortunately, the orbital and spin parameters derived by \citet{cam03}, \citet{kir04} and \citet{fal05a} for XTE J1807-294 form {\it XMM-Newton} observations were based on the presumption that the orbital period is fixed to the preliminary value proposed by \citet{mar03b} because the time base line is too short to constrain the orbital period.  On the other hand, the $\sim$150d's {\it RXTE} observations allow further binary parameter refinement, that is essential for further investigation.  In this paper, we report the pulsation analyses of XTE J1807-294 using archived {\it RXTE} data (section~\ref{dar}).  Due to the complexity of pulse phase variation, a trend removal technique is employed to reveal precise orbital parameters for the system.  Pulse phases (binary barycenter-corrected) are found to exhibit anomalous negative shifts during the soft flares in all energy bands between 2-20 keV.  The evolution of the non-flare pulse phases shows that the neutron star was spun-up during the outburst. Significant $\sim500$ $\mu s$ soft lags for the non-flare pulses is detected at energies from 2 to 20 keV.  Analysis results, including possible implications for coincident soft flares and anomalous negative phase shifts are discussed in section \ref{dis}.

\section{{\it RXTE} Observations} \label{ob}

Extensive follow-up observations for XTE J1807-294 were conducted by {\it RXTE} soon after the source was discovered until it went into its quiescent state.  We analyzed the available archived {\it RXTE} data observed from February 27 to July 29, 2003. The data used to study the timing properties of XTE J1807-294 were collected by {\it RXTE} PCA \citep{jah96}, which consists of five gas-filled Proportional Counter Units (PCUs) with the total collecting area $\sim$6500 $cm^2$ and sensitive in the photon energy range 2 to 60 keV.  The PCA data of all the observations were collected in the GoodXenon mode with a time resolution of $\sim$1 $\mu$s. In addition, to compare the pulsation behavior and light curve (see section~\ref{ppfc}), the data from both PCA and High Energy X-ray Timing Experiment (HEXTE) were used for spectral fitting to obtain the fluxes for the all available XTE J1807-294 {\it RXTE} observations.

\section{Data Analysis and Results} \label{dar}

\subsection{Binary Orbital Parameters} \label{bop}
First, all the events collected by two Event Analyzers (EAs) were combined; then, the event arrival times were further corrected to the barycenter of the solar system, using JPL DE200 ephemeris and the source position determined by {\it Chandra} observation \citep{mar03b}.  Only the events detected by the top layer of each PCU and with photon energies between 2 to 10 keV were selected in order to improve the sensitivity.  Subsequently, all the selected events were divided into 150s data segments for further analysis.

The epoch folding search technique with a search resolution of $4 \times 10^{-9}$ sec was applied to test the significance of pulsation for each 150s data segment and to obtain the preliminary binary orbital parameters.  The test results show that the 190 Hz pulsation can be detected only in the first $\sim$90d (up to TJD 12783) out of the overall $\sim$150d's observations.  In order to correctly estimate the orbital parameters, only those data segments with the maximum $\chi ^2$ value from epoch folding search larger than three times of the degrees of freedom (dof=31 for 32 bins/period, false alarm probability $< 10^{-4}$) were taken into account. The trial period with maximum $\chi ^2$ was considered the best pulsar period for each data segment, and the corresponding error was evaluated using the equation proposed by \citet{lea87}.  The Doppler effect owing to the orbital motion can be clearly observed with the best-fitted mean pulsar frequency $\nu=$190.62353(3) Hz (or $P_s=5.2459421(9)$ ms) and binary orbital parameters: $P_{orb}=$40.0735(2) min, ${\it a_x}\sin {\it i}=4.83(6)$ lt-ms and $T_{\pi /2}=$TJD12720.68245(6). Neither orbital eccentricity nor the spin frequency derivative can be significantly detected.  Improved orbital parameters and the evolution of the pulsar spin frequency can be obtained from the pulse arrival time delay; hence the above results were applied as the initial guess values for the following phase analysis.

\begin{figure}
\begin{center}
\epsscale{.80}
\includegraphics[angle=90,scale=.80]{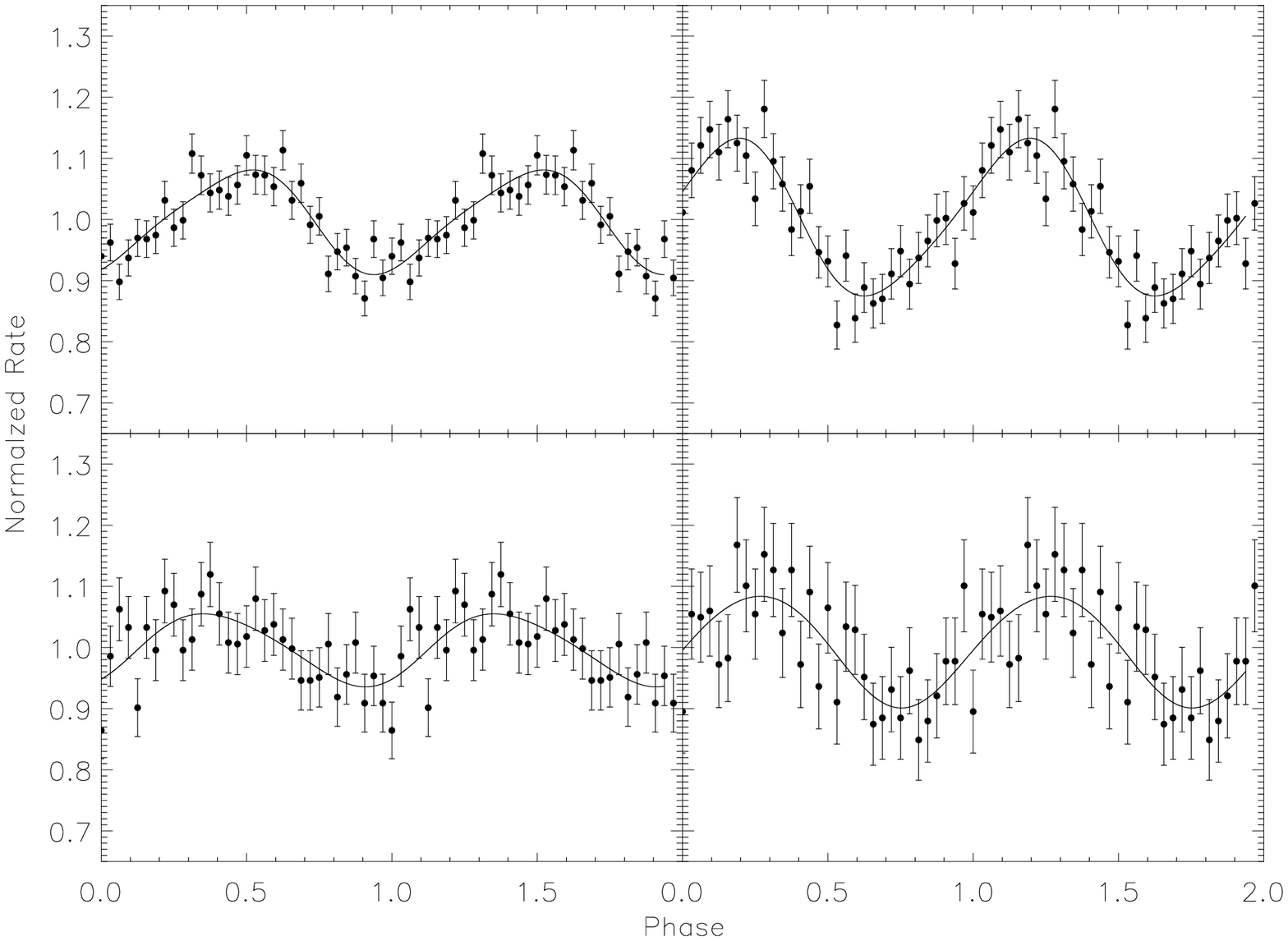}
\end{center}
\caption{The pulse profiles (2-10 keV) of XTE J1807-294 measured on TJD12697 (upper left), TJD12712 (upper right), TJD12720 (lower left) and TJD12775 (lower right). The solid line on each plot is the pulse profile fitted by the two-Fourier component model.  \label{pulse_profile}}
\end{figure}

Irrespective of whether the complex spin change could be a result of the hot spot moving on the surface of the neutron star or accretion torque, we can expect the intrinsic phase (after binary barycenter correction) to be very complicated. Therefore, we employed an alternative method to obtain the precise orbital parameters instead of using a method analogous to the one proposed by~\citet{dee81} which is applied to accretion millisecond pulsars (e.g.~\citealt{bur06}).  Because the radial velocity amplitude of the neutron star ($K$) in XTE J1807-294 is small ($K/c \sim 2 \times 10^{-6}$), the pulsation can be still detected in a 150-sec data segment if the folding frequency is close to the true spin frequency of the neutron star.  The pulse arrival time delay due to the orbital motion can be obtained from the pulse phase shift.  The pulse profile folded with the mean pulsar period derived from the orbital Doppler effect is better fitted with a two-Fourier component model (fundamental + first overtone).  Typical pulse profiles are shown in Figure~\ref{pulse_profile}.  The peak of the fitted profile was chosen as the fiducial point for the following analysis.  The error of the phase for each 150s data segment was estimated from 1000 runs of Monte Carlo simulations.  In addition to the time delay due to the orbital motion, significant linear phase drift (more than 1 cycle per day) was detected in $\sim$3.5 hours' exposure because of slight inconsistency between the folding period and the neutron star spin period. To minimize the possible cycle count ambiguity, we adjusted the folding period to $P_{fold}=5.245942726$ ms (or $\nu_{fold}=$190.6235070 Hz), based on the best pulsar ephemeris derived by the observation on TJD12700, and applied it to the entire data.  To test if the intrinsic phase evolution is indeed complex, we fitted the observed pulse phase variation with the combined function of the time delay due to binary orbit plus evolution of intrinsic pulse phase modeled as a polynomial, that is

\begin{equation}\label{e_phi}
\phi_{s}^{int}(t)=\phi_{s}^{int}(T_0)+(\nu_{fold}-\nu)(t-T_0)- {1 \over 2} \dot{\nu}(t-T_0)^2-{1 \over 6} \ddot{\nu}(t-T_0)^3 + ......
\end{equation}

\noindent However, no simple polynomial (up to 8th order) can yield acceptable fit (reduced $\chi^2 \approx$ 4.2). We conclude that the evolution of intrinsic pulse phase is too complicated to be modeled as a simple polynomial.

\begin{figure}
\begin{center}
\includegraphics[angle=90,scale=.80]{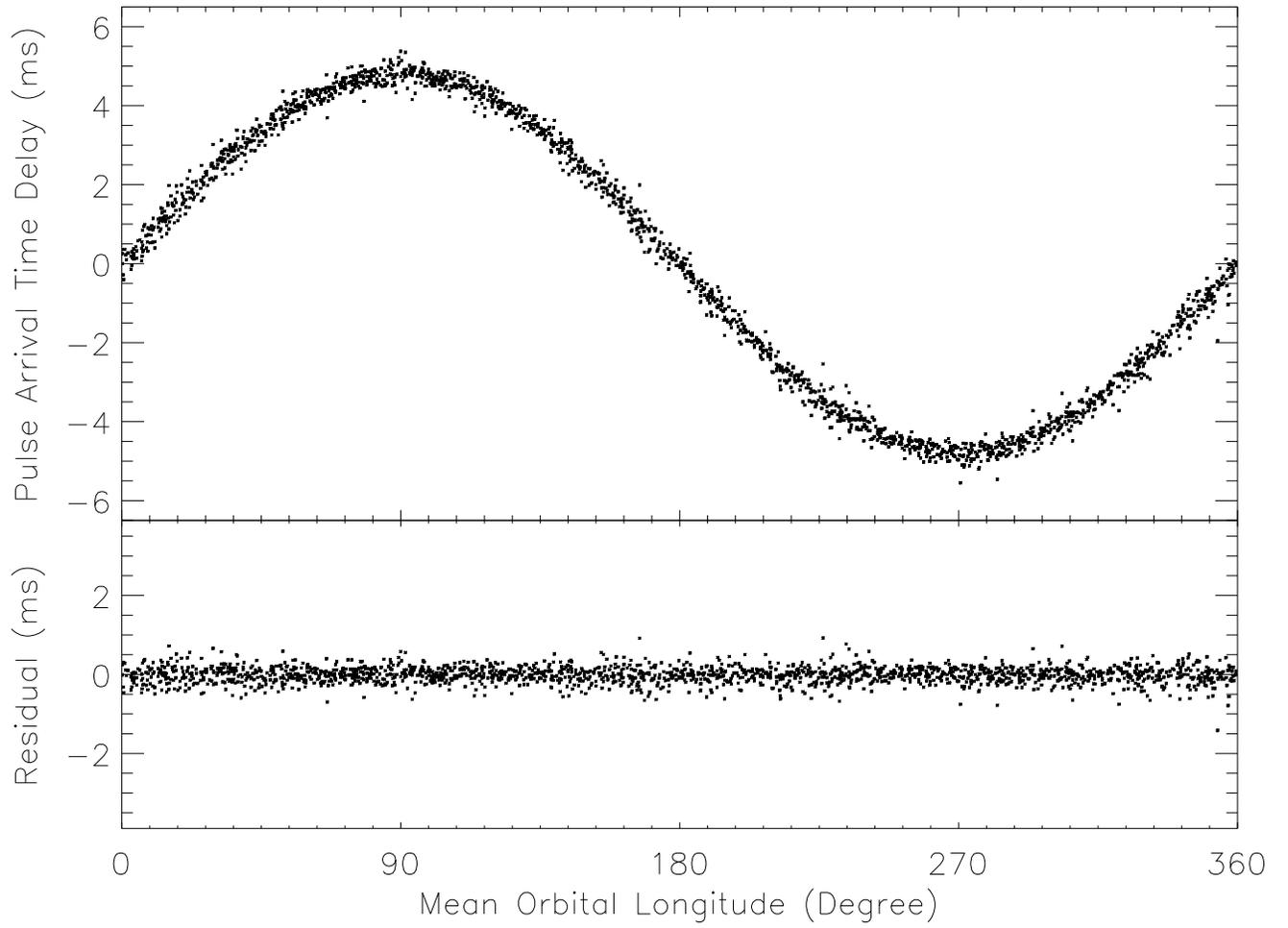}
\end{center}
\caption{Pulse arrival time delay (after the phase trend removal) due to the orbital motion (top) and the fit residuals (bottom).\label{orb_delay}}
\end{figure}

Since further study of the nature of accreting millisecond pulsar requires precise binary orbital ephemeris, we alternatively tried to remove the intrinsic phase trend for daily observation.  Although the evolution of intrinsic pulse phase was complicated for the first $\sim$90d's of observations, the higher order terms of the intrinsic phase drift can be neglected for the daily observation because of the short exposure ($\la$3.5 hrs).  Therefore, we fitted a linear trend plus a single sinusoidal function due to the orbital motion with a fixed trial orbital period derived from the Doppler effect and then removed only the linear trend from the phases.  The trend-removed phases were fitted with a circular orbit model to obtain updated orbital parameters. To further improve the orbital parameters, the new orbital period was used as the trial orbital period and above processes were repeated. The iterations of this procedure continued until improvements of all parameters were at least an order smaller than the corresponding errors.  The final trend-removed phases are well fitted with a circular orbit whose best parameters are listed in Table~\ref{tb_op}. The F-test shows that adding the eccentricity to the orbital model dose not significantly improve the fit. The F value is 1.3, which indicates that probability for improvement of the fit occurring by chance can be as high as $\sim$30\%.   The pulse arrival time delay due to the binary orbital motion is shown in Figure~\ref{orb_delay}.

\subsection{Pulse and Flux Correlation}\label{ppfc}

\begin{figure}
\begin{center}
\includegraphics[angle=90,scale=.80]{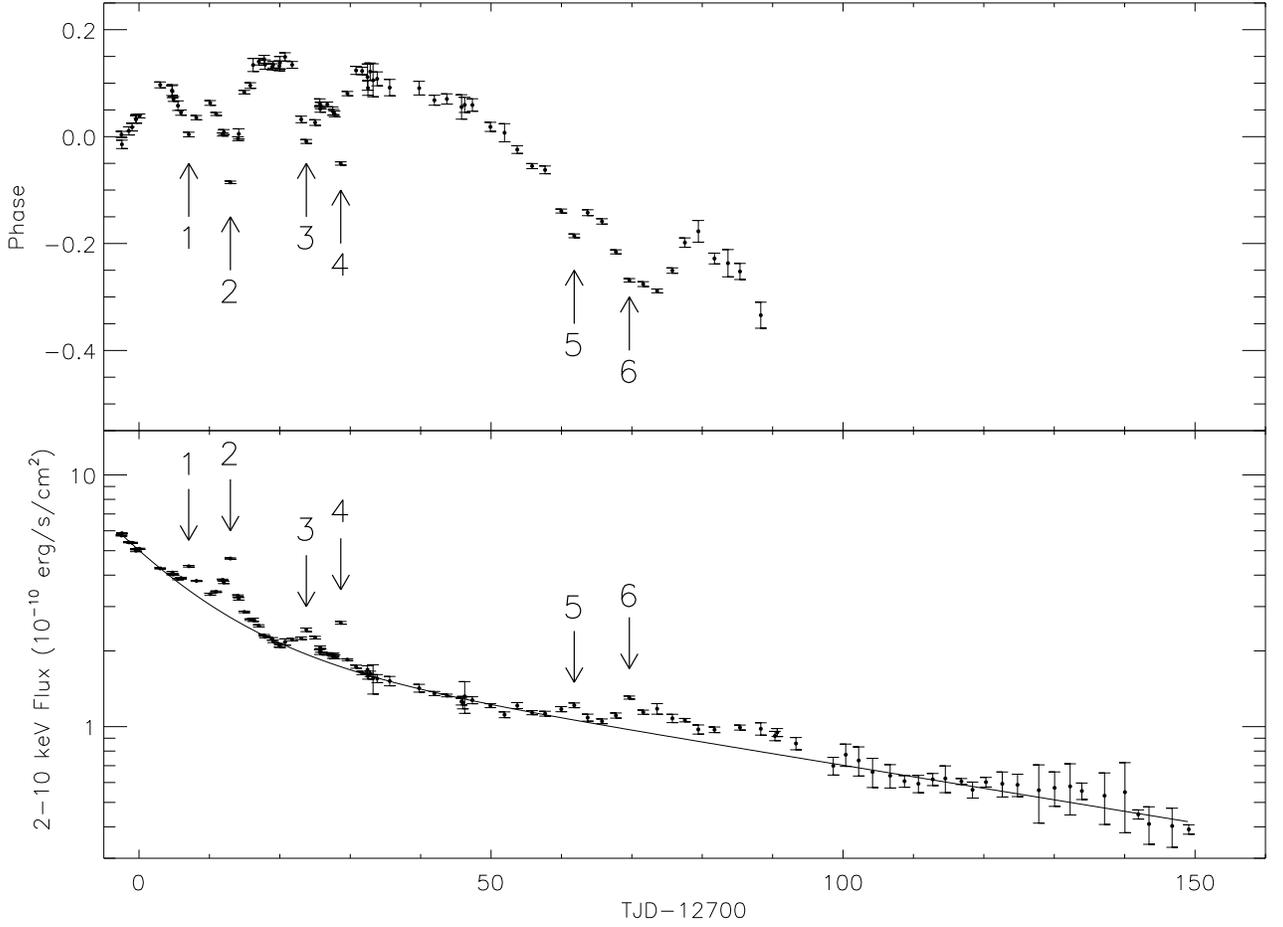}
\end{center}
\caption{Comparison between evolution of 2-10 keV pulse phase folded with a constant frequency plus Keplerian orbit (i.e. Eq~\ref{fold}) (top panel) and 2-10 keV flux (bottom panel).  A total six flares were detected during the flux declination.  The arrows in both panels indicate the flare peak time.  Clear flare and phase shift coincidence can be seen. The solid line on the bottom plot represents the best fitted two-component exponential model for the non-flare state fluxes.\label{time_vs_phase_flux}}
\end{figure}

\begin{figure}
\begin{center}
\includegraphics[angle=90,scale=.80]{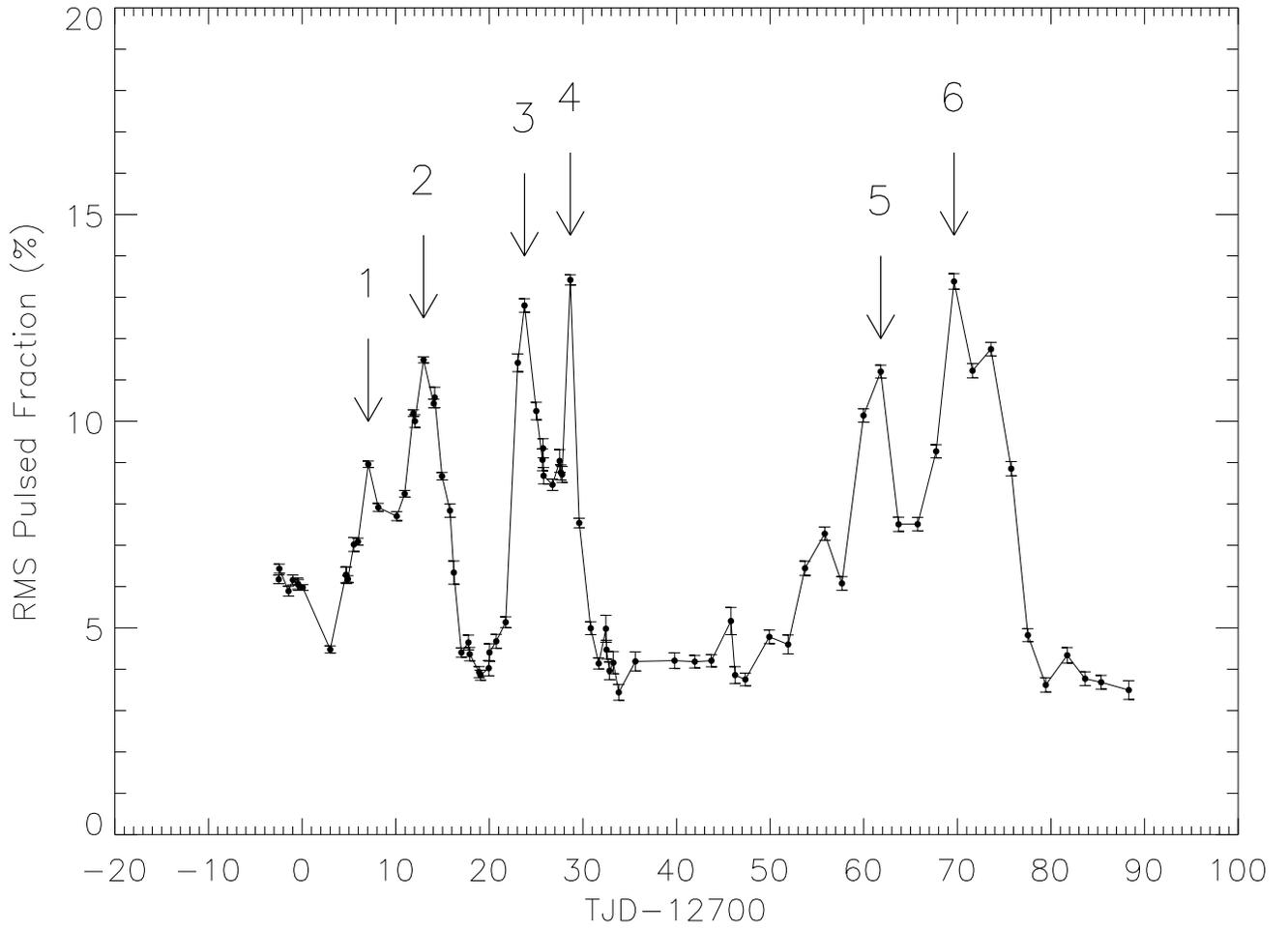}
\end{center}
\caption{RMS pulsed fraction vs. time.  Similar to Figure~\ref{time_vs_phase_flux}, the arrows indicate the times for the six flare peaks, showing the pulse amplitude and flare coincidence. \label{pulse_fraction}}
\end{figure}

The pulse phase or spin frequency variation of the neutron star can be studies on basis of precise orbital parameters.  For obtaining the binary barycenter-corrected pulse phases (hereafter, pulse phases), we folded the event time $t_i$ with a time-variable frequency due to the orbital Doppler effect using the parameters ($P_{orb}$, $a_x \sin i$, and $T_{\pi/2}$) derived in section~\ref{bop} as

\begin{eqnarray}\label{fold}
\phi_i & = &frac \left\{ \int_{T_0}^{t_i}\nu_0+\nu_0{{2\pi a_x \sin i} \over {cP_{orb}}}\sin \left[{{2 \pi (t-T_{\pi / 2})} \over {P_{orb}}}\right] dt \right\} \\
\nonumber & = & frac \left\{\nu_0(t_i-T_0)-\nu_0{{a_x \sin i} \over {c}}\cos \left[{{2 \pi (t_i-T_{\pi / 2})} \over {P_{orb}}}\right]+\nu_0{{a_x \sin i} \over {c}}\cos \left[{{2 \pi (T_0-T_{\pi / 2})} \over {P_{orb}}}\right]\right\} 
\end{eqnarray}

\noindent where $frac$ is the fraction of cycle counts calculated inside the braces, $\nu_0$ is folded frequency (see section~\ref{bop}) and $T_0$ is the phase zero epoch of the pulsation.  This operation is equivalent to first correcting the event times to the binary system barycenter and then folding with a constant frequency $\nu_0$.  The pulse profile for each observation ID was constructed by through binning the event phases.  Figure~\ref{time_vs_phase_flux} shows the pulse phase variation during the 2003 outburst.  The pulse phases change like a smooth curve with several anomalous negative phase shifts occurring around March 9 (TJD 12707), 15 (TJD 12713), 26 (TJD 12724), 30 (TJD 12728) and also around May 2(TJD 12761) with time scale of several days but become somewhat irregular after May 11 (TJD 12770) (see top panel of Figure~\ref{time_vs_phase_flux}).  The first four anomalous phase shifts are coincident with the ``puny'' flares discovered by \citet{zha06} (see Figure 1 of their paper).   \citet{zha06} also reported that the pulsed fraction increases with the flares.  Therefore, we examined the fractional root mean square (RMS) amplitude for the first $\sim$90d's observation. Figure~\ref{pulse_fraction} shows two additional significant enhancements in pulsed fraction around TJD 12761 and 12769.  The former one is coincident with the fifth anomalous negative phase shift.

To further investigate the coincidence of anomalous phase shifts and flares, we made a 2-10 keV light curve for $\sim$150d observations through spectral fitting.  We used the archived PCA standard-2 and HEXTE data and adopted the spectral model of absorbed blackbody plus a power law to obtain the 2-10 keV fluxes. This model gives reasonable fit results for all observations with reduced $\chi^2$ of 0.6-1.6. The bolometric correction estimated by the mean ratio of the integrated 2-200 keV to 2-10 keV is $2.41 \pm 0.27$.   Figure~\ref{time_vs_phase_flux} shows the 2-10 keV flux of XTE J1807-294 during 2003 outburst.  Four strong flares with peak fluxes detected on TJD 12707, 12713, 12724 and 12728, as well as two weak flares with peak fluxes detected on TJD12761 and 12769, can be clearly seen on the light curve.  We labeled these flares from 1 to 6 according to their chronological order.  Pulse phases exhibit negative phase shifts, and the pulsed fractions increase during the flare states. From Figure~\ref{time_vs_phase_flux} and Figure~\ref{pulse_fraction}, the flares are likely to appear in pairs with peak separations of 5.9d for flares 1 and 2, 4.8d for flares 3 and 4 and 7.8d for flares 5 and 6.

\begin{figure}
\begin{center}
\includegraphics[angle=90,scale=.80]{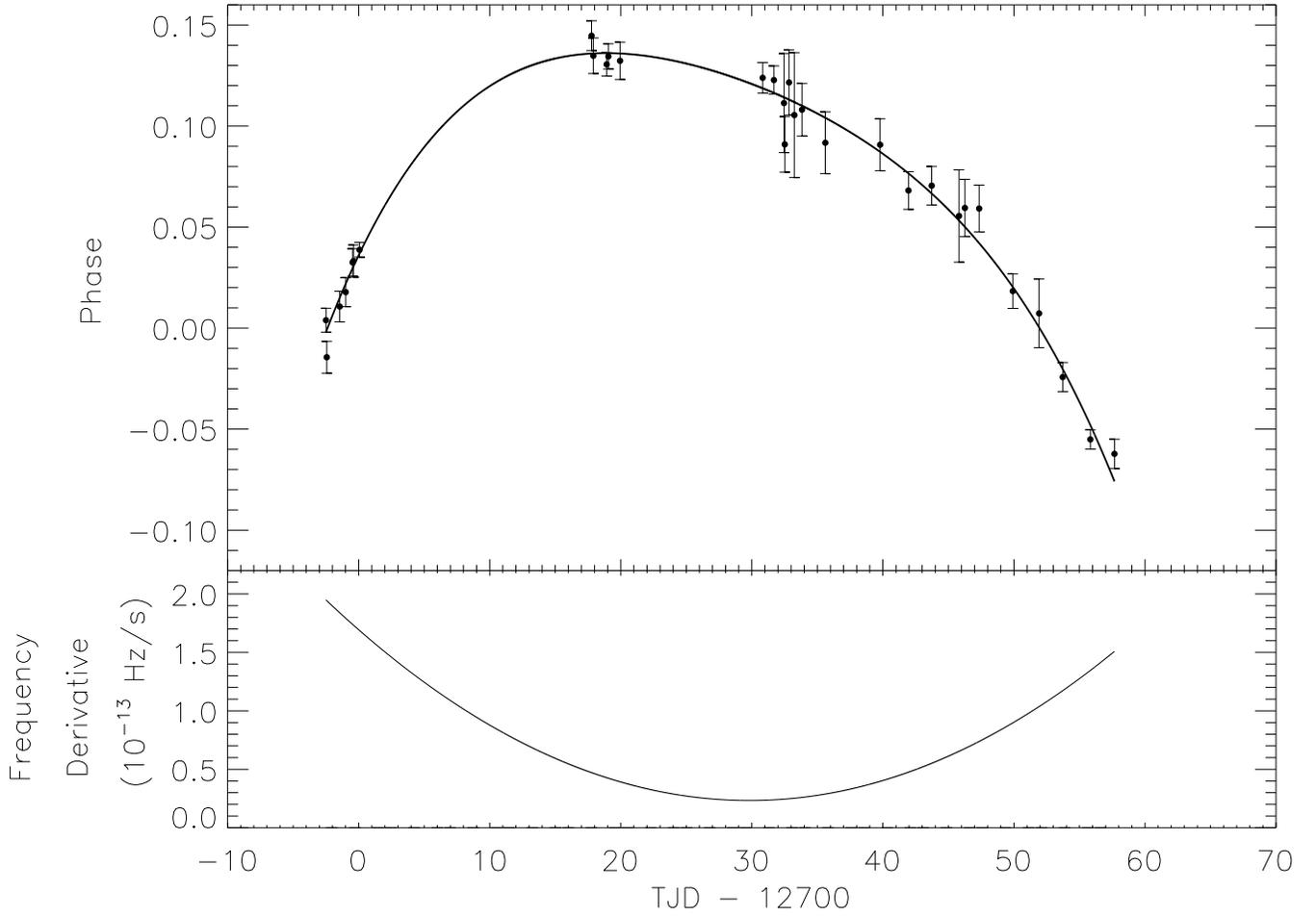}
\end{center}
\caption{Top: Phase evolution of non-flare state from TJD12697 to 12757.   The solid line is the best-fit 4th order polynomial.  Bottom: The spin frequency derivative ($dot \nu$) derived from the best-fit parameters from phase evolution, assuming that the hot spot is fixed on the surface of neutron star. \label{phase_ev}}
\end{figure}

These negative phase shifts are more likely due to the motion of the accretion footprint rather than the spin-down torque acting on the neutron star, which will be discussed in section~\ref{dis_ps}.  Further analysis results for the pulse behaviors during the flare states are presented in section~\ref{edpb}.  On the other hand, the smooth pulse phase evolution for the non-flare state during the first $\sim$60d of the outburst is possible owing to the accretion torque.  We defined the epochs for the flare states: TJD 12703 to 12716 for flares 1 and 2, TJD 12720 to 12730 for flares 3 and 4 and TJD 12759 to 12778 for flares 5 and 6.  To investigate the smooth pulse phase evolution during the non-flare states, we removed the phases during the flare states and also discarded the ones after TJD 12778.  The non-flare phases were fitted with a polynomial, starting from a quadratic function and then subsequently adding higher order terms.  The F-test was applied to verified if the additional higher order term significantly improves the fitting (with confidence level $>$ 90\%).  We found that the phase evolution for the non-flare states is better described by a 4th order polynomial (see Figure~\ref{phase_ev}).  On the other hand, imperfect solar system barycenter correction due to source position error results in correction error for event arrival time \citep{lyn98}, which further leads to systematic errors in spin parameters. From Eq. 5.3 in  \citet{lyn98}, the systematic errors of the pulse phase and spin parameters, including phase zero epoch ($T_0$), spin frequency ($\nu$) and its derivatives ($d^n \nu / dt^n$)  can be expressed as

\begin{mathletters}
\begin{eqnarray}\label{syserr}
\delta \phi & \approx & {{R_E \nu \delta \theta}\over {c}} \sin[\Omega_E(t-T_\gamma)-\lambda-\beta]  \\
\delta T_0 & = & {{\delta \phi} \over \nu} \approx {{R_E  \delta \theta}\over {c}} \sin[\Omega_E(t-T_\gamma)-\lambda-\beta]  \\
\delta \nu & = & {{d (\delta \phi)} \over dt}\approx \Omega {{R_E \nu \delta \theta}\over {c}} \cos[\Omega_E(t-T_\gamma)-\lambda-\beta]  \\ 
\delta \Bigl({{d^n \nu} \over {dt^n}}\Bigr)& = &{{d^{n+1} (\delta \phi)} \over dt^{n+1}}
\end{eqnarray}
\end{mathletters}

\noindent where $R_E$ is the orbital radius of the Earth, $\Omega_E$ is Earth's orbital angular velocity, and $T_\gamma$ is the time of vernal point.  The best fitted spin parameters and the corresponding statistic and systematic errors are listed in Table~\ref {tb_sp}.  The frequency derivative $\dot \nu(t)$ derived from the best fitted spin parameters shows that the neutron star was spun-up during the first $\sim$60d of the 2003 outburst (see Figure~\ref{phase_ev}).   Further, the non-flare light curve can be well described as the combination of a fast and a slow exponential decay component with decay time constants of $\tau_{f}=11.4 \pm 0.2$ d and $\tau_{s}=95 \pm 2$ d respectively (see Figure~\ref{time_vs_phase_flux}).

\subsection{Energy-dependent Pulse Behaviors}\label{edpb}

\subsubsection{Soft Lags for Non-flare Pulses}\label{sl}

\citet{cui98} discovered that a high-energy pulse leads a low-energy pulse by about 200 $\mu$s (referred to as ``soft lags'') from 2 to 10 keV but saturates at a higher energy for the accreting millisecond pulsar SAX J1808.8-3658.   This energy-dependent pulse arrival time has been well explained by the scenario proposed by \citet{gie02} and \citet{pou03}.  The emissions mainly consist of a soft blackbody component from a heated spot on the neutron star surface and a hard component from the Comptonization of the blackbody photons (seed photons) in the plasma heated by an accretion shock slab with Thomson optical depth of 0.3 to 1.  The hard Comptonization component leads the soft blackbody component as a result of the different angular distribution of the soft and hard components and the Doppler boosting effect.  The low-energy pulses lag behind the high-energy pulses because the latter contain relative less soft blackbody component.  For the photons energies greater than $\sim$10 keV, only the Comptonization component is left so the hard leads saturate.  The emission spectra of many accretion-powered millisecond pulsars are well described by this two-component model, including XTE J1807-294 \citep{fal05a}.  Subsequent to the discovery of soft lags in SAX J1808.8-3658, similar phenomenon was also seen in some of accreting millisecond pulsars, such as XTE J1715-305 \citep{gie05} and HETE J1900.1-2455 \citep{gal07}.  However, for IGR J00291+5934, the soft lags increase up to 7 keV and then decrease at higher energy bands instead of saturation \citep{fal05b}.  The decrease at higher energy bands is likely due to the time lags caused by Compton upscattering \citep{sun80,vau97}.

\begin{figure}
\begin{center}
\includegraphics[angle=90,scale=.80]{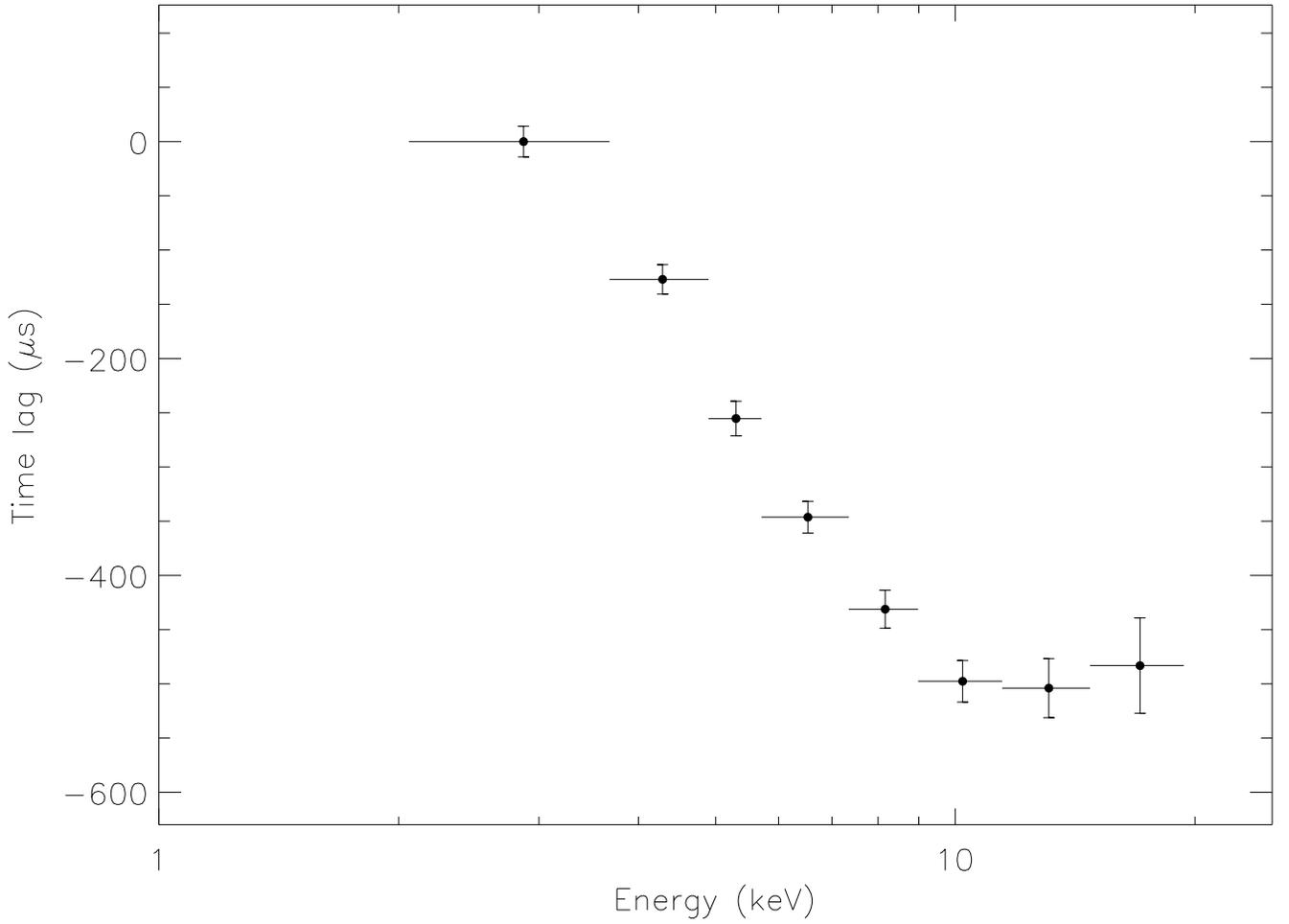}
\end{center}
\caption{Energy-dependent pulse arrival time delay relative to the softest energy band (2-3.7 keV) for the non-flare pulse. Negative value indicates that the pulse leads the softest one.  The leading saturates at $\sim$10 keV about 500 $\mu$s ($\sim$0.1 cycle). \label{soft_lag}}
\end{figure}

\begin{figure}
\begin{center}
\includegraphics[angle=90,scale=.80]{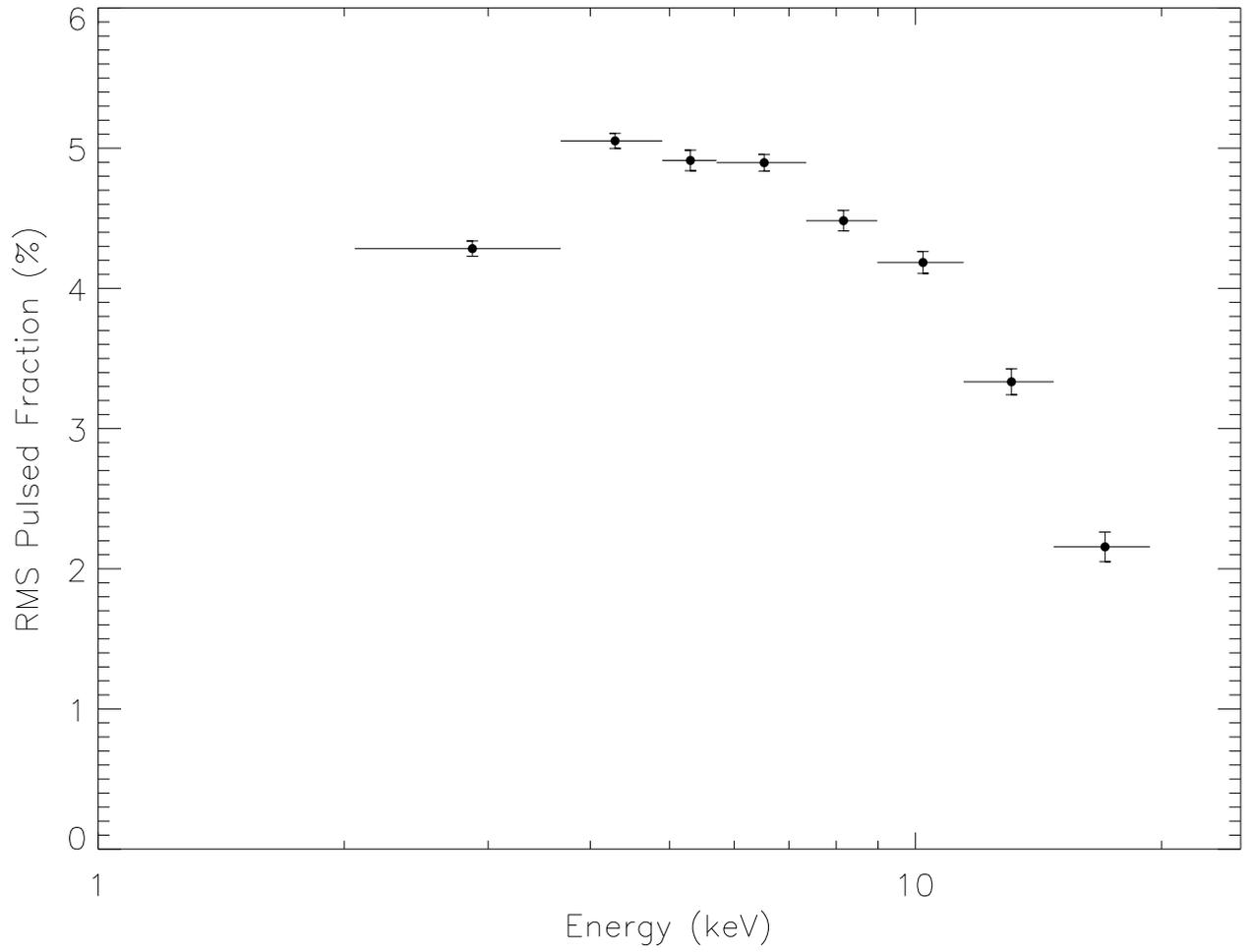}
\end{center}
\caption{Energy-dependent mean pulsed fraction for the non-flare state.\label{pulse_frac_band}}
\end{figure}

To investigate the energy-dependent pulse behaviors for XTE J1807-294, we divided 2-20 keV photons into 8 energy bands for the first $\sim$60d observations.  Since we assumed hot spot to be fixed during the non-flare state, we folded the non-flare event times of each energy band with a ephemeris (hereafter, the best ephemeris) from the variable observed pulse frequencies due to the orbital Doppler effect and neutron star spin parameters from the 4th order polynomial fitting (see Table~\ref{tb_sp}).  Thus, pulse phases of the non-flare state are expected to be a constant with time. The average pulse profile for the non-flare state of each energy band was obtained by binning the corresponding event phases.  The energy-dependent pulse arrival time delays, measured relative to the pulsation of the softest band (2-3.6 keV), were obtained through the cross correlation of the pulse profiles.  The high energy pulses lead soft pulses up to $\sim 500$ $\mu s$ ($\sim 0.1$ cycle) from 2 to 10 keV and the leads saturate for 10 to 20 keV pulse (Figure~\ref{soft_lag}), similar to that seen in SAX J1808.8-3658 \citep{cui98}. On the other hand, the fractional RMS pulse amplitude slightly increases from 4.5\% to 5\% in the 2-4 keV range, remains constant to 7.5 keV, and largely decreases to only 2\% at 20 keV (see Figure~\ref{pulse_frac_band}).

\subsubsection{Phase Shifts during Flare States}\label{ps}

\begin{figure}
\begin{center}
\includegraphics[angle=90,scale=.80]{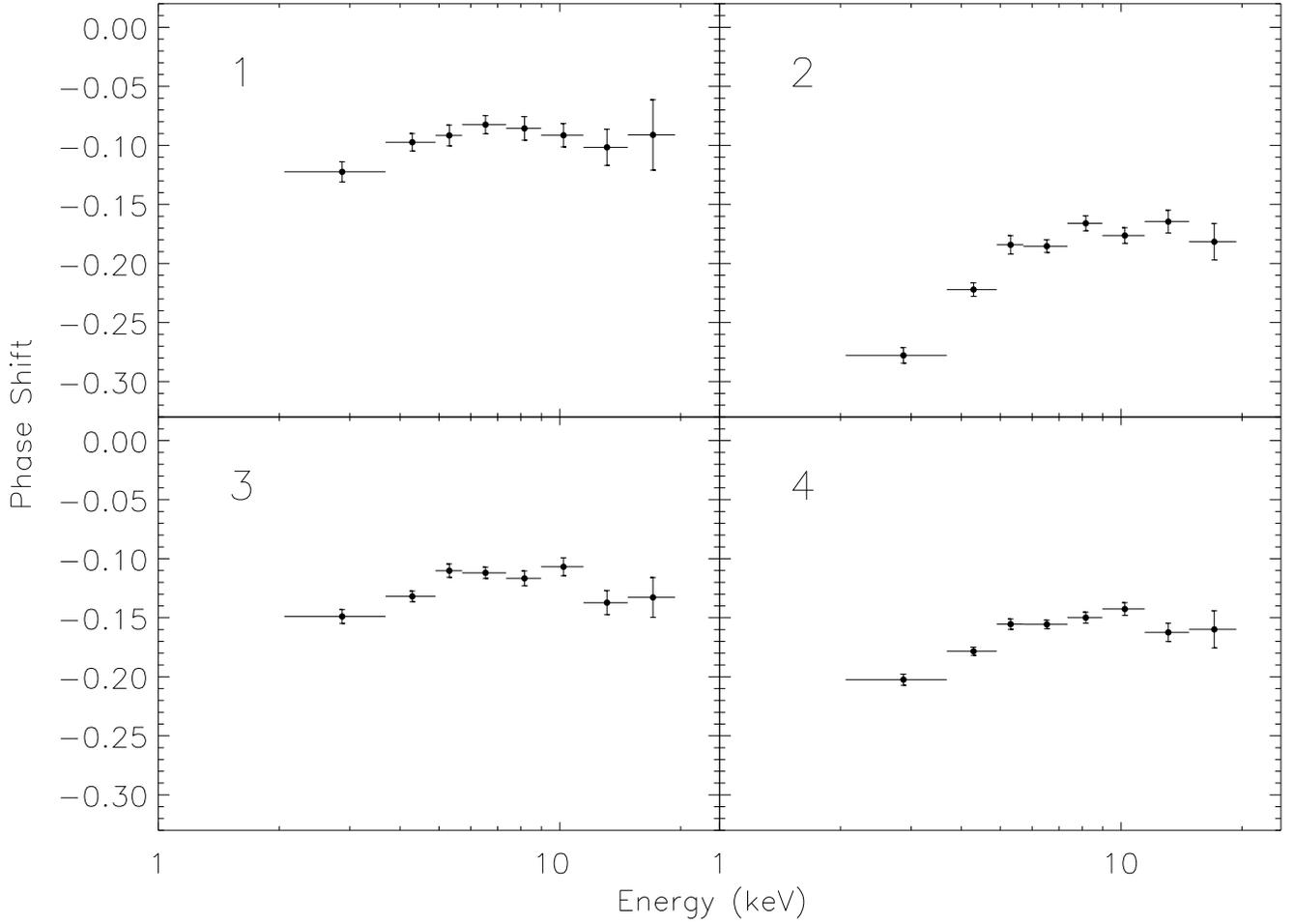}
\end{center}
\caption{Phase shifts of different energy bands for the flare states measured on the first four flare peak times.  The flare index is labeled on the up-left corner of each plot.  Significant phase shifts can be seen in higher energy bands, although the lowest energy band has the largest phase shift \label{phase_shift_band}}
\end{figure}

Significant phase shift, and its correlation to the flux and fractional pulse amplitude, has never observed for other accreting millisecond pulsars.  To further understand its origin, we studied the energy-dependent phase shifts during the flare state.  The pulse phase defined here is identical to the one in section~\ref{sl}, that is, folded with the best ephemeris.  The phase shift for each energy band, defined as the phase difference between the flare state and the corresponding mean phase of the non-flare state, was obtained through the cross correlation method.  However, since this ephemeris is applicable for the first $\sim$60d data, only the phase shifts of the first four flares can be well-defined.  Figure~\ref{phase_shift_band} shows that the phase shifts of different energy bands when the fluxes of flares attained their maximum values for the flares 1 to 4.  Significant phase shifts can be observed in all energy bands for all the four flares.  In general, the phase shifts decrease for higher energy bands from 2 to 5 keV, almost a constant between 5-11 keV and slightly increase for the hardest energy band.

\section{Discussion}\label{dis}

\subsection{Orbital Parameters}\label{dis_op}

We have derived the precise binary orbital parameters for the X-ray binary XTE J1807-294. The orbital parameters have been previously reported by \citet{mar03a}, \citet{cam03}, \citet{kir04}, and \citet{fal05a}, whose values are listed in Table~\ref{tb_opc} for comparison.  All the parameters are consistent with each other except the epoch of the mean longitudes for the binary orbit (i.e. phase zero epochs).  The epoch of mean longitude is chosen as when mean longitude=$90^\circ$ ($T_{\pi / 2}$) in this paper.  \citet{fal05a} claimed that they used same reference point, but it deviates from ours by about $\sim$0.5 phase.  We examined their arguments and found that they  probably used the positive-crossing time in time vs. pulsar period plot (see Figure 6 in \citet{fal05a}), which is equivalent to the mean longitude=$270^\circ$, as the phase zero epoch.  If their reference time is indeed $T_{3\pi / 2}$ instead of $T_{\pi / 2}$ , the value proposed by them and the one in this paper are consistent.  On the other hand, the epoch of mean longitude considered by \citet{kir04} is equivalent to the mean longitude=$0^\circ$, which is close but still significantly different from what we found (more than 8$\sigma$).  Their value, and the one proposed by \citet{cam03}, is probably due to a short observation time span and, as mentioned in \citet{fal05a}, the complex method used by these authors.

\subsection{Phase Shift during Flare State}\label{dis_ps}

With the precise orbital parameters, we have found significant pulse phase shifts on the time scale of one to several days coincident with the flares as well as pulsed fraction enhancements.  If the hot spot is assumed to be fixed on the neutron star surface and the pulse phase variation truly reflects the history of the change in the spin frequency of the neutron star, fitting a quadratic curve for the pulse phases around the flare states implied a spin-down rates $\sim 10^{-11}$ Hz/s.  It is much larger than the theoretical expectation value ($\sim10^{-13}$ Hz/s, see \citealt{rap04}) or the reported values for accretion-powered millisecond pulsars ($\sim 10^{-12}-10^{-13}$ Hz/s, see e.g. \citealt{bur07} ).  In addition, if the flares are caused by an enhanced accretion, the additional accretion torque should further spin-up the neutron star rather than spin it down \citep{gos79,rap04}.   Thus, the flare-coincident negative phase shifts are highly unlikely due to the accretion torque.

Another possible explanation for the anomalous pulse phase shifts is the hot spot moving on the surface of neutron star.  Assume that the neutron star is rotating eastward as the Earth and that the hot spot is fixed at a position in the northern hemisphere during the non-flare state.  As the accretion rate increases in a flare state, the hot spot moves eastward, in the same direction as that of the neutron star rotation, so that the pulse peak arrives to an observer earlier than the one expected for the non-flare state and results negative phase shift. At the time when the flare reaches its maximum, this movement can be as large as about 70$^\circ$ ($\sim$0.2 cycle) in longitude.  When the flare's flux decreases, the hot spot moves westward and eventually returns to its original location at the end of the flare.

The correlation between phase shift and flux has been suggested for other accreting millisecond pulsars, although the no flare similar to that observed in XTE J1807-294 has ever been detected in other accreting millisecond pulsars. \citet{papi07} reported the discovery of high anti-correlation between X-ray flux and the pulse phase jitter ($\sim \pm0.05$) around the mean quadratic phase trend in the accreting millisecond pulsar XTE J1814-338.  Such a correlation is very similar to what we found in XTE J1807-294 during the flare states although the correlation between the pulsed fraction and the flux is still unknown for XTE J1814-338.  Recently,~\citet{bur06} discovered an anomalous large phase jump for SAX 1808.4-3658 during its 2002 outburst, which is also likely as a result of a large accretion rate change (see Figure 1 in \citealt{bur06}). 

\citet{zha06} suggested that the flares seen in XTE J1807-294 could result from accreting blobs of high mass density, lasting for hours to days, at the inner edge of the accretion disk.  The positive correlation of kHz QPO frequency and pulsed fraction implies that the surface emission properties of neutron star are very sensitive to the inhomogeneous disk flow \citep{zha06}.  We found that not only the pulsed emissions increase but also the position of the hot spot moving on neutron star surface.  Since the longitudinal movement of the hot spot  on the neutron star results in the phase shift, it is also possible that the hot spot moves in latitude, which could be a part of the reason for the pulse amplitude enhancements during the flare states.

The flux and phase shift coincidence has also been seen in accreting pulsars with much lower spin frequency.  \citet{stri96} reported that during the bursting X-ray pulsar GRO J1744-28 ($\nu = 2.1$Hz) type II bursts, likely a consequence of accretion instability, the pulse arrival time lags by $\approx 90$ ms relative to the non-burst state. Pulsed fraction is also strongly enhanced during the bursts \citep{kou96}.  \citet{mil96} proposed a scenario to describe the pulse phase shifts.  Because of the misalignment of the neutron star spin axis and magnetic dipole moment, the accretion flow is asymmetric and its footprint forms an arc surrounding the magnetic pole.  The enhanced accretion causes the footprint to move around magnetic pole as a consequence of the deformation of the field lines.  The azimuth movement of the footprint with respect to the spin axis results in the observed phase shift. Except that the phase shift is negative for XTE J1807-294 but positive (pulse arrival delay) for GRO J1744-28 and the different time scales for flare (several days) and type II burst (several seconds), the coincidence of phase shift, pulse amplitude enhancement and burst in GRO J1744-28 is very similar to what we report for XTE J1807-294.  The scenario suggested by \citet{mil96} can probably give us a clue to explain the observed behaviors of XTE J1807-294.

Form the discovery of the coincidence between the flare and phase shift in XTE J1807-294, as well as the correlation between the X-ray flux and the pulse phase jitters in XTE J1814-338 \citep{papi07}, hot spot position on accreting millisecond pulsar is sensitive to accretion flow, as suggested by~\citet{zha06}.  Although we assume that the accretion hot spot is still fixed on the surface neutron star during the non-flare state, we cannot exclude the possibility that the hot spot marginally drifts as the X-ray flux changes.  Therefore, it is difficult to determine whether the pulse phase variation is a consequence of the spin frequency change or hot spot drift on the neutron star surface.  If the correlation between the hot spot drift and accretion flow is applicable to all accreting millisecond pulsars, the measurements of the spin parameters (e.g. $\dot \nu$ ) from which accretion torque is estimated, are not reliable, including the ones listed in Table~\ref{tb_sp} for XTE J1807-294.  This may also explain the large disparity between the measured spin frequency derivative and the theoretical value calculated form the accretion torque reported by~\citet{mor03} for SAX J1808.4-3658 in its 2002 outburst.

\section{Conclusion}\label{con}

\paragraph{1.} Due to complicated intrinsic phase variations, to obtain the precise orbital parameter, we folded our data sets with a constant frequency and used the trend removal technique to remove the local linear pulse phase trend from the observed pulse phase variation for daily observations. We obtained more precise orbital parameters consistent with the ones previously reported (except for $T_{\pi /2}$).

\paragraph{2.} The binary barycenter-corrected pulse phases change smoothly except for several anomalous negative phase shifts coincident with flux flares and pulse amplitude enhancements.  The pulse fractional amplitude variation indicates that there were six flares during the first $\sim$90d of the XTE J1807-294 in its 2003 outburst.  The phase changes of non-flare pulse imply that the neutron star was spun-up during the first 60d of the outburst under the assumption that the hot spot was fixed on the surface of the neutron star during the non-flare states.  

\paragraph{3.} The analysis results of the energy-dependent pulse arrival time indicates that hard pulses lead soft pulses up to $\sim 500$ $\mu s$ from 2 to 10 keV, and the lead saturates from 10 to 20 keV, which agrees with the scenario proposed by \citet{gie02} and \citet{pou03}.  

\paragraph{4.} Our discovery of the coincidence of flare and phase shift, as well as its correlation with pulse amplitude and KHz QPO~\citep{zha06} in XTE J1807-294, implies that probably due to weak magnetic confinement on the accreting material, the hot spot position on the accreting millisecond pulsars is sensitive to the accretion flow.   

\acknowledgments

The authors thank R. E. Taam, J. H. Swank, H. C. Spruit, M. C. Miller and D. Lai for helpful discussions.  This research has made use of data obtained through the High Energy Astrophysics Science Archive Research Center (HEASARC), provided by the NASA's Goddard Space Flight Center.  This work is supported by the grants from National Science Council NSC 94-2112-M-008-003 and NSC 95-2112-M-008-026-MY2.



\clearpage

\clearpage

\clearpage

\clearpage

\clearpage

\clearpage

\clearpage

\clearpage

\clearpage

\begin{table}
\begin{center}
\caption{Orbital Parameters for XTE J1807-294 from {\it RXTE} Observation\label{tb_op}}
\begin{tabular}{cc}
\\
\tableline\tableline
Parameter & Value \\
\tableline
Orbital period, $P_{orb}$ (min) \dotfill & 40.073601(8)\\
Projected semimajor axis, ${\it a_x} \sin {\it i}$ (lt-ms) \dotfill & 4.823(5)\\
Epoch of 90$^\circ$ mean longitude, T$_{\pi/2}$ (TJD/TDB)\dots \dots  & 12,720.682574(6)\\
Mass function ($M_{\odot}$) \dotfill &$1.497(4) \times 10^{-7}$ \\
Eccentricity, $e$ \dotfill & $<$0.006(2$\sigma$)\\
$\chi^2$/Degree of freedom  \dotfill & 2171/1880\\
\tableline
\end{tabular}
\end{center}
\end{table}

\clearpage


\begin{table}
\begin{center}
\caption{Neutron Star Spin Parameters for XTE J1807-294\label{tb_sp}}
\begin{tabular}{cccc}
\\
\tableline\tableline
Parameter & Value \tablenotemark{a}& Statistic & Systematic\\
          &                        & Error     & Error \tablenotemark{b}\\
\tableline
Phase zero epoch, $T_0$ (TJD/TDB) \dotfill & 12700.00000000(2) &$1.0\times 10^{-10}$ &$1.7\times 10^{-8}$\\
Frequency at $T_0$, $\nu_0$ (Hz)\dotfill & 190.62350686(8) &$1.4\times 10^{-8}$\ & $7.4\times 10^{-8}$\\
$d\nu/dt$ at $T_0$ (Hz/s) \dotfill & $1.7(3) \times 10^{-13}$ &$3.0\times 10^{-14}$ & $1.1\times 10^{-14}$\\
$d^2\nu/dt^2$ at $T_0$ (Hz/s$^2$) \dotfill & $-1.1(1) \times 10^{-19}$ & $1.0\times 10^{-20}$&$2.9\times 10^{-21}$\\
$d^3\nu/dt^3$ (Hz/s$^3$) \dotfill & $4.4(4) \times 10^{-26}$ & $4.0\times 10^{-27}$&$4.3\times 10^{-28}$\\
$\chi^2$/Degree of freedom  \dotfill & 22.76/26\\
\tableline
\end{tabular}
\tablenotetext{a}{The parameter errors are evaluated from quadratic sum of statistical and systematic errors}
\tablenotetext{b}{The systematic errors are estimated from the $\sim1^{\arcsec}$ source position error (see  ~\ref{ppfc})}

\end{center}
\end{table}

\clearpage

\begin{table}
\begin{center}
\caption{Measured Orbital Parameters for XTE J1807-294\label{tb_opc}}
\begin{tabular}{lllllc}
\\
\tableline\tableline
 Mission & $P_{orb}$ & {\it a} sin{\it i} & Epoch of mean& Phase for epoch & Reference\\
& (min) & (lt-ms) &  longitude(TJD)& of mean longitude\tablenotemark{a}& \\

\tableline
{\it RXTE}&40.0741(5)& - & - & - & 1\\
{\it XMM-Newton}&40.0741(fixed)&$4.7^{+1.2}_{-0.04}$&12720.68644(13)& 0.140(5)& 2\\
{\it XMM-Netwon}&40.0741(fixed)&$4.8(1)$&12720.67415(16)&0.699(6) & 3\\
{\it XMM-Netwon}&40.0741(fixed)&$4.75(39)$&12720.668(2)&0.48(7) & 4\\
{\it RXTE}&40.073601(8)&$4.823(5)$&12720.682574(6)& -& 5\\

\tableline
\end{tabular}
\tablenotetext{a}{Folded with the orbital ephemeris proposed in this paper.  See ~\ref{dis_op} for the discussions about the inconsistencies}
\tablerefs{(1) \citet{mar03a}; (2) \citet{cam03}; (3) \citet{kir04}; (4) \citet{fal05a}; (5) this work}
\end{center}
\end{table}

\end{document}